# Interfacial Properties of Bilayer and Trilayer Graphene on Metal Substrates

Jiaxin Zheng,[1,2,†] Yangyang Wang,[1,†] Lu Wang,[3] Ruge Quhe,[1,2] Zeyuan Ni,[1] Wai-Ning Mei,[3] Zhengxiang Gao,[1] Dapeng Yu,[1] Junjie Shi,[1] and Jing Lu[1,*]

[1]State Key Laboratory for Mesoscopic Physics and Department of Physics, Peking University, Beijing 100871, P. R. China

[2]Academy for Advanced Interdisciplinary Studies, Peking University, Beijing 100871, P. R. China

[3]Department of Physics, University of Nebraska at Omaha, Omaha, Nebraska 68182-0266

[†]These authors contributed equally to this work.

*Corresponding author: jinglu@pku.edu.cn

One popular approach to prepare graphene is to grow them on transition metal substrates via chemical vapor deposition. By using the density functional theory with dispersion correction, we systematically investigate for the first time the interfacial properties of bilayer (BLG) and trilayer graphene (TLG) on metal substrates. Three categories of interfacial structures are revealed. The adsorption of B(T)LG on Al, Ag, Cu, Au, and Pt substrates is a weak physisorption, but a band gap can be opened. The adsorption of B(T)LG on Ti, Ni, and Co substrates is a strong chemisorption, and a stacking-insensitive band gap is opened for the two uncontacted layers of TLG. The adsorption of B(T)LG on Pd substrate is a weaker chemisorption, with a band gap opened for the uncontacted layers. This fundamental study also helps for B(T)LG device study due to inevitable graphene/metal contact.

**Introduction**

Graphene has become a 'hot topic' due to its extraordinary properties[1-3] and wide range of possible applications[4-8]. Synthesis of high-quality graphene on a large scale is the foundation for its application. Among different preparation methods, growing graphene on transition metals including Cu[9-14], Co[15], Ni[16,17], Pt[18], Pd[19], Au[20], Ru[21,22], Rh[23], and Ir[24,25] via chemical vapor deposition (CVD) is overwhelming because of high-quality, low preparation temperature, scalable production, and easy transfer to other substrates. Through CVD method, not only single layer graphene (SLG) but also few-layer graphene can be synthesized[13,14,16]. Among few-layer graphene, bilayer (BLG) and trilayer graphene (TLG) are the most extensively studied materials, partially due to the fact that there is an electrically tunable band gap in BLG[26-30] and ABC-stacked TLG[31-34] and meanwhile the carrier mobility is not degraded, which are critical for their application in transistor. Additionally, in an actual device, graphene has to be contacted with metal electrode. Therefore, the interfacial properties of B(T)LG and metal contacts should be clarified.

The interfacial properties between SLG and metals have been systematically studied[23,35-38]. The adsorption of SLG on Al, Ag, Cu, Au, and Pt (111) surfaces is a weak physisorption, which preserves the Dirac cone of SLG. By contrast, the adsorption of SLG on Ti (0001) surface, and Ni, Co, and Pd (111) surfaces is a strong chemisorption, which perturbs the electronic structure of SLG significantly. SLG is $n$-type doped by Al, Ag, Cu, Ti, Co, Ni and Pd, but $p$-type doped by Au and Pt. However, a systematic study on the interfacial properties between B(T)LG and metal substrates is lacking and leaves three fundamental issues open: (1) How do the B(T)LG/metal interfacial properties change with the species of metals? In view of the additional layer and easier break of inverse symmetry, new features may emerge when BLG and ABC-stacked TLG are contacted with metal substrates compared with SLG cases. (2) How do the TLG/metal interfacial properties depend on the stacking style of TLG? The second issue becomes especially crucial in light of the fact that the ABC- and ABA-stacked TLG possess inversion and mirror symmetries, respectively, resulting in a distinct response to electric field: A vertical electric field can open a band gap in ABC-stacked TLG but increase the overlap between the conduction and valence bands in ABA-stacked TLG instead[31-34,39]. (3) Previous theoretical studies have been reported that the contact effects between SLG and Al electrodes can affect the transport properties of SLG devices significantly by inducing an extra conductance minimum at the Dirac point of the contacted region and giving rise to an electron-hole asymmetry[40]. It is open how the metallic contacts affect the transport properties of B(T)LG devices.

In this Article, we provide the first systematic investigation on the interfacial properties of BLG and TLG on a variety of metals (Al, Ag, Cu, Au, Pt, Ti, Co, Ni, and Pd) by using the density functional theory (DFT) with dispersion correction and establish the general physical picture of the B(T)LG/metal interfaces. Three categories of B(T)LG/metal interfacial structures are revealed in terms of the adsorption strength and electronic properties: The adsorption of B(T)LG on metal substrates (Al, Ag, Cu, Au, and Pt) is a weak physisorption in the first category of interfaces, but a band gap can be opened and its size depends on the possessed symmetries in graphene. The adsorption of B(T)LG on metal substrates (Ti, Ni, and Co) is a strong chemisorption in the second category of interfaces, and bands of the upper layer graphene of BLG are intact while a stacking-insensitive band gap is opened for the two uncontacted layers of TLG. The adsorption of B(T)LG on metal substrates (Pd) in the third category of interfaces is a weaker chemisorption, with a stacking-sensitive band gap opened for the two uncontacted layers of TLG and a band gap of 0.12 eV opened for the upper layer graphene of BLG. Finally, we design a two-probe model made of BLG contacted with Al and Ti electrodes, respectively, and calculate their transport properties by using *ab initio* quantum transport theory. Distinct transport properties are observed: A clear conductance gap rather than a conductance minimum appears at the Dirac point of the contacted region with Al as electrodes but this gap is full filled with Ti as electrodes.

## Results
### Geometry and stability of B(T)LG on metal substrates

The most stable configurations of the SLG/metal interfaces are shown in Figure 1a (metal = Co, Ni, and Cu, named after top-fcc interface) and 1b (metal = Al, Ag, Pt, Au, and Ti)[35,36]. We choose the two configurations as the initial configuration of B(T)LG/metal contacts. As shown in Figure 1c and d, TLG/metal contacts favor split alignment of the first graphene layer with respect to metals compared with those of the S(B)LG/metal contacts.

The calculated key data are presented in Table 1. The binding energy $E_b$ of the B(T)LG/metal contact is defined as

$$E_b = (E_G + E_M - E_{G/M})/N \tag{1}$$

Where $E_G$, $E_M$, and $E_{G/M}$ are the relaxed energy for B(T)LG, the clean metal surface, and the combined system, respectively, and $N$ is the number of carbon atoms in a unit cell. The interfacial distance $d_{C-M}$ is defined as the average distance of innermost graphene to metal surfaces. The B(T)LG/metal contacts can be classified into three categories according to the binding strength and the interfacial distance. In the first category of interfaces (Al, Ag, Cu, Au, and Pt (111) substrates), B(T)LG are physisorbed on these metal substrates with smaller binding energies of $E_b = 0.032 - 0.063$ eV and larger interfacial distances of $d_{C-M} = 3.13 - 3.53$ Å; for TLG both quantities are insensitive to the stacking order. With larger $E_b = 0.094 - 0.210$ eV and smaller $d_{C-M} = 2.04 - 2.34$ Å, B(T)LG are strongly chemisorbed on Ti (0001) surface, and Ni and Co (111) surfaces, forming the second category of interfaces. Differently, the binding in the second category of interfaces is always stronger by $0.02 - 0.03$ eV for the ABC stacking style compared with the ABA stacking style. The adsorption of the third category of interfaces (Pd substrate) is a weak chemisorption (or strong physisorption), which is intermediate between the physisorption and strong chemisorption, with $E_b = 0.08$, 0.085, and 0.103 eV and $d_{C-M} = 2.70$, 2.54, and 2.50 Å for BLG, ABA-, and ABC-stacked TLG, respectively. The same classification is applicable to the SLG/metal contacts[35,36], and thus the graphene layer number has little effect on the adsorption categories. The flat planes of BLG and TLG are all kept in the first category of interfaces. But the innermost graphene layer buckles slightly with buckling heights of $0.01 - 0.11$ Å in the second category of interfaces and $0.01 - 0.02$ Å in the third category. The buckling height difference also reflects the difference of interaction strength among three categories of interfaces.

**Electronic structure of BLG on metal substrates**

The classification is also in accordance with the electronic structure of B(T)LG on metal surfaces. We calculate the band structures of the first category of interfacial structures. As showed in Figure 2, the band structure of the BLG can be clearly identified in these systems because of the weak interaction. Two important changes in the BLG bands are noteworthy:

One is the Fermi level ($E_f$) of BLG is shifted upward or downward when contacted with the first class of metal surfaces, similar to SLG cases [35,36]. BLG is *n*-type (upward shift) doped when contacted with Ag, Al, and Cu but *p*-type doped (downward shift) when contacted with Au and Pt. This phenomenon can be attributed to the different work functions of BLG ($W_G$, the calculated value is 4.58 eV)) and metal surface ($W_M$). The Fermi level shift is defined as $\Delta E_f = E_f - E_D$, where $E_D$ is the middle energy of the band gap of the BLG adsorbed on metal substrates. The Fermi level shift $\Delta E_f$ as a function of ($W_M - W_G$) is plotted in Figure 3(a). The change tendency of $\Delta E_f$ with ($W_M - W_G$) is in accordance with that of the SLG cases[35,36]. The crossover point from *n*- to *p*-type doping is not at $W_M - W_G = 0$ but at about $W_M - W_G = 0.4$ eV (the LDA result for the SLG cases is $W_M - W_G = 0.9$ eV) [35,36]. At the crossover point, there is no charge transfer between metal and BLG. Therefore, the value of $W_M - W_G$ at that point reflects the potential step resulting from the BLG-metal chemical interaction ($\Delta_c = 0.9$ eV). Such a chemical interaction effectively reduces $W_M$ by $\Delta_c$. As a result, a larger $W_M$ is needed to induce *p*-type doping in both BLG and SLG.

The other feature of the electronic structures of BLG in the first class of interfaces is the appearance of a band gap of $E_g = 0.102 - 0.200$ eV (Table 1), which is absent in their SLG counterparts. These band gaps are smaller than the maximum band gap of 0.25 eV opened in BLG under a vertical electric field[29,41] and the maximum band gap of 0.34 eV opened in SLG sandwiched between hexagonal boron nitride under a vertical electric field[42]. The mechanism of band gap opening can be explained by a BLG/metal contact model, as shown in Figure 3(b). We use $\Delta n$, $\Delta n_1$, and $\Delta n_2$ to denote the transferred charge on metal surfaces, the bottom layer graphene, and the upper layer graphene, respectively. The electron transfer assumedly creates a uniform electric field $E$ and $E_1$ between the sheets. The potential difference between the two graphene sheets is

$$\Delta U = U_2 - U_1 = -\alpha \Delta n_2 - \Delta_c, \ \alpha = e^2 d_0/\varepsilon_0 \kappa \qquad (2)$$

where $\varepsilon_0$ is the dielectric constant of vacuum and $\kappa$ the relative dielectric constant of graphene. $\Delta U$ is thus proportional to the transferred charge on the upper layer graphene. Due to $\Delta U \neq 0$, the inversion symmetry of A-B stacked BLG is broken. As a result, a band gap is induced, which has been confirmed by the tight-binding calculations in the system of depositing potassium on BLG[43] and in few layer graphene under a vertical electric field[32,44,45]. The change of the band gap $E_g$ as a function of $\Delta E_f$ is shown in Figure 3(c). $E_g$ increases with the increasing $|\Delta E_f|$ in both the *n*- and *p*-type doping regions. The cause lies in the fact with the increasing doping level in the *n*-type doping region (reflected by $|\Delta E_f|$), the more charge is transferred, and $|\Delta n_2|$ and $|-\alpha\Delta n_2 - \Delta_c|$ gets larger, leading to a larger $|\Delta U|$ and thus a larger $E_g$. The $E_g - \Delta E_f$ data in the *n*-type doping region even can be roughly fitted by a linear function $E_g = -0.42 \times \Delta E_f + 0.05$ eV (black dashed line). It implies that there is a band gap of 0.05 eV for BLG physisorbed on metal substrates due to $\Delta_c$ even if the doping level is zero.

Experimentally, the current on/off ratio of a BLG field effect transistor (FET) is significantly improved by one order of magnitude when the channel BLG is deposited by Al, suggestive of opening of a transport gap in BLG[46]. This result is in agreement with our calculation that a band gap is opened for BLG on Al substrate. Furthermore, in terms of our calculations, the current on/off ratio of BLG FETs can also be improved by deposition of Cu, Ag, Au, and Pt on channel BLG.

The band structures of the second category of interfaces are shown in Figure 4(a-e). The bands of the bottom layer graphene are strongly disturbed and hybridized with the *d* bands of the metal surfaces, and the characteristic cone at the *K* point is destroyed. The minority spin bands of Ni and Co hybridize with both the $\pi$ and $\pi^*$ bands of the bottom layer graphene, whereas the majority spin bands of Ni and Co hybridize chiefly with the $\pi$ bands of the bottom layer graphene, because some of the minority spin bands are above the Dirac point of graphene while most of the majority spin bands are below the Dirac point. The 3*d* bands of Ti hybridize with both the $\pi$ and $\pi^*$ bands of the bottom layer graphene because they are distributed widely both below and above the Dirac point. The hybridization between Ti and the bottom graphene is so strong that we even can't identify the bands of the bottom layer graphene, a result consist with the largest binding energy of BLG on Ti. By contrast, the band structure of the upper layer graphene is almost intact and can be clearly identified, and the Dirac cone at the *K* point is preserved perfectly. The only change is that $E_f$ of the upper layer graphene is shifted upwards, induced an *n*-type doping. The Fermi-level shift of the upper layer graphene is −0.364/−0.280 eV (−0.428/−0.381 eV) for the bands of the majority/minority spin of Ni (Co), respectively, and −0.283 eV for the bands of Ti contact (Table 1). All these changes indicate that the bottom layer graphene has been strongly bonded with the metal surfaces and formed a new surface. The interaction between the upper layer graphene and this new surface is a physisorption, so the electronic structure of the upper layer graphene is well preserved. Our band structure of BLG on Ni (111) surface is in agreement with that reported by Gong et al.[47] Another theoretical work shows that BLG is chemisorbed on Ru (0001) surface, with an intact band structure for the upper layer graphene and a strongly perturbed band structure for the bottom layer graphene[22]; apparently, BLG/Ru (0001) also belongs to the second class of interfaces.

The Fermi-level shift of the upper layer graphene on Ni (Co) substrate is −0.364/−0.280 eV (−0.428/−0.381 eV) for the majority/minority spin bands, respectively (Table 1), which is comparable with that (−0.361 eV) of BLG on Al substrate. But the work function of Co/Ni is 4.97/4.93 eV, and $W_{Co/Ni} - W_G = 0.390/0.350$ eV, very close to the crossing point $W_M - W_G = 0.4$ eV. It appears that the Fermi-level shift should be very small for Co (Ni) contact according to our previous model. Nevertheless, the bottom layer graphene is strongly chemisorbed on the Co (Ni) surface, and they actually form a new surface. The work function of the new surface is $W_{new} = 4.33/4.32$ eV (Co/Ni), and one has $W_{new} - W_G = -0.250/-0.260$ (Co/Ni). The Fermi-level shift of the upper graphene on Ni/Co substrate is $\Delta E_f \sim 0.280$ eV from Figure 3(a), which well accounts for the actual values (−0.364/−0.280 eV for Ni and −0.428/−0.381 eV for Co)

In Figure 4(f), we show the band structure of the third category of interfaces (BLG/Pd). The conduction bands of the bottom layer graphene can be identified but it is hard to identify the valence bands, indicating that the Pd 4d/5s bands are chiefly hybridized with the valence bands of the bottom layer graphene because most of the Pd 4d/5s bands are below the Dirac point of BLG. Similar to the second class of interfaces, the band structure of the upper layer graphene can be clearly identified, but a band gap of 0.124 eV is opened. $E_f$ of the upper layer is shifted upward by 0.160 eV, indicative of *n*-type doping. This unique interfacial electronic structure of BLG/Pd is ascribed to the fact that the intermediate interaction between BLG and Pd surface preserves the partial electronic properties of the bottom layer and it does not overwhelm the intrinsic graphene interlayer coupling. The dipole field induced by Pd-graphene charge transfer breaks the inverse symmetry of the two graphene layers, and a band gap is opened and is chiefly reflected in the upper layer graphene. The $E_g - \Delta E_f$ datum of BLG on Pd substrate falls in the $E_g - \Delta E_f$ fitting curve for BLG physisorbed on metal substrates in the *n*-doping region (Figure 3c).

**Electronic structure of TLG on metal substrates**

The electronic structures of the first category of TLG/metal interfaces are plotted in Figure 5. It is clearly shown that the band structures of both ABA- and ABC-stacked TLG are preserved, accompanied by a Fermi level shift and an opened band gap. The same as BLG cases, TLG is doped with electrons by Al, Ag, and Cu contacts and with holes by Au and Pt contacts. Table 1 summarizes the evolution of the Fermi-level shift $\Delta E_f = E_D - E_f$ and band gap $E_g$ of TLG. The Fermi-level shift $\Delta E_f$ as a function of $W_M - W_G$ is shown in Figure 6a for ABA- and ABC-stacked TLG ($W_G$ is the calculated work function of TLG, 4.52 eV). The crossover point from *n*- to *p*-type doping is about $W_M - W_G = 0.56$ for ABA-stacked TLG and 0.55 eV for ABC-stacked TLG.

Both the Fermi-level shift $\Delta E_f$ and work function difference $W_M - W_G$ are less sensitive to the stacking order. However, the stacking order affects significantly the band gap of TLG: ABA-stacked TLG has generally smaller band gaps of $E_g = 0 - 0.061$ eV while ABC-stacked TLG has generally larger band gaps of $E_g = 0 - 0.249$ eV. The current on/off ratio of TLG FET is expected to be increased by deposition of Al, Cu, and Ag on channel TLG due to a band gap opening. As shown in Figure 6b and Figure S1, the sizes of the band gap of both ABC- and ABA-stacked TLG rough linearly depend on $\Delta E_f$. The band gaps of ABC-stacked TLG are apparently electron-hole asymmetric: they are significantly larger in the *n*-type doping region ($E_g = 0.181 - 0.249$ eV) than those in the *p*-type doping region ($E_g = 0 - 0.018$ eV) at the same $|\Delta E_f|$. The band gaps of ABA-stacked TLG are slightly electron-hole asymmetric: the band gaps are slightly larger in the *n*-type doping region ($E_g = 0.032 - 0.061$ eV) than those in the *p*-type doping region ($E_g = 0$ eV) given the same $|\Delta E_f|$ (Figure S1). The mechanisms of band gap opening in ABA- and ABC-stacked TLG and the reason of electron-hole asymmetry in band gaps can also be explained with a TLG/metal contact model (Figure S2).

The electronic structures of the second category of interfaces (TLG/Ti, Co, and Ni contacts) are plotted in Figure 7. Similar to BLG cases, both the $\pi$ and $\pi^*$ states of the innermost graphene layer are strongly hybridized with the 3$d$ states of Ti, and minority-spin 3$d$ states of Co and Ni. Only the $\pi$ states of the innermost graphene layer are strongly hybridized with the Co and Ni majority-spin 3$d$ states. The strongly coupled innermost graphene layer serves as an active buffer and effectively passivates the metal $d$ states at the interface. As a result, the electronic structure of the two uncontacted layers is nearly intact except that a band gap is opened, similar to the first category of interfaces for BLG/metal. The band gaps of the uncontaced BLG are less sensitive to the stacking mode, with $E_g = 0.100 - 0.229$ eV, which approach the maximum band gap of freestanding BLG obtained from the theoretical $(0.25 - 0.28$ eV$)^{28,32}$ and experimental $(0.25$ eV$)^{29}$ studies. The band gap opening is attributed to a potential energy difference between the two uncontacted graphene layers, which is $\Delta U_{32} = -\alpha \Delta n_3 - \Delta_c'$, where $\Delta_c'$ is the potential step resulting from the interaction between the second graphene layer and the chemically bonded innermost graphene-metal system. Such a potential energy difference breaks the inversion symmetry of the two uncontacted graphene layers, thus opening a band gap. TLG is $n$-type doped in both stacking styles for the work function of the uncontaced BLG $W_G$ is larger than that of the new graphene-metal surface $W_{new}$.

    The electronic structures of the third category of interfaces (TLG/Pd contact) are shown in the second and fourth top panels of Figure 7. TLG is $n$-type doped by Pd substrate in both stacking styles. Though the $\pi$ states of the innermost graphene layer are perturbed strongly, the $\pi^*$ states are only slightly affected, leading to BLG-like valence bands (two bands visible near the $K$ point) and TLG-like conduction bands (three bands visible near the $K$ point). The cause is the same as BLG cases: most of the Pd 4$d$ states are below the Dirac point of graphene and they can only hybridize with the $\pi$ states of the innermost graphene layer. Analogous to the physisorption cases, the band gap of the two uncontaced layers of TLG on Pd substrate is strongly dependent on the stacking mode, with $E_g = 0.064$ and $0.308$ eV for ABA and ABC stacking styles, respectively. The latter band value is even marginally larger than the maximum band gap of freestanding BLG under a uniform electric field obtained from the theoretical $(0.25 - 0.28$ eV$)^{28,32}$ and experimental $(0.25$ eV$)^{29}$ studies and comparable with the maximum band gap of SLG sandwiched between $h$-BN sheet under a uniform electric field[42]. The unique behavior of TLG/Pd is also agreement with the intermediate

binding between typical physisorption and chemisorption.

**Transport properties of BLG contacted with metal electrodes**

Finally, we further study how the interfacial properties affect the transport properties of BLG devices when contacted with metallic Al and Ti leads. The two-probe model is presented in Figure 8(a), and the distance between the Al/Ti lead and BLG is 3.45/2.18 Å, according to our DFT results. The transmission spectrum of the device with Al electrodes is shown in Figure 8(b), where a minimum ($D_{ch}$) close to $E_f$ due to the Dirac point of the channel BLG and a 0.22eV gap at $E - E_f = -0.6$ eV are observed. By contrast, there are only one transmission minimum close to $E_f$ for pure BLG without metal electrode (Inset in Figure 8(b)) and two transmission minima for SLG contacted with Al electrodes (one close to $E_f$, and the other at $E - E_f = -0.6$ eV due to the Dirac point of SLG in the lead) (Figure 8(c)). The transport gap in Figure 8(b) originates from a gap of the same size in the projected density of states (PDOS) of the BLG in the lead, as shown in Figure 8(d), because the transmission coefficient of the device, $T(E)$, is connected with the PDOS of the channel and the two electrodes *via* the relation[48]:

$$T(E) \propto \frac{g_{ch}(E)g_L(E)g_R(E)}{g_{ch}(E)g_L(E) + g_{ch}(E)g_R(E) + g_L(E)g_R(E)} \tag{3}$$

where $g_{ch}(E)$ and $g_{L/R}(E)$ are the PDOS of the channel and the left/right lead, respectively. Both gaps can be attributed to the band gap of $E_g = 0.20$ eV of corresponding infinite BLG contacted with Al electrodes. The transmission spectrum with Ti electrodes is shown in Figure 8(e). Compared with Al contacts, the transmission minimum due to the Dirac point of the channel BLG remains but the transmission gap is absent because the characteristic conical point at the $K$ point of the bottom layer graphene is destroyed. There is neither gap nor Dirac point in the PDOS of BLG in the lead (Figure S3). From the transmission spectra, Ti electrode can transport a larger current than Al electrode. In the light of similar interfacial properties of TLG/metals to BLG cases, the same contact effect on transport properties of TLG devices as BLG devices with Al and Ti electrodes is expected: Ti electrode can transport a larger current than Al electrode. The difference in the transport properties between Al and Ti electrodes in BLG is also reflected from a difference of the transmission eigenchannel at $E - E_f = -0.6$ eV and at the ($\pi/3a$, 0) point of the *k*-space. As displayed in Figure 8(f), the transmission eigenvalue at this point nearly vanishes with Al electrodes, and the incoming wave function is nearly completely scattered and unable to reach to the other lead. By contrast, the transmission eigenvalue at the point is 0.96 with Ti electrodes, and the incoming wave function is scattered little and most of the incoming wave is able to reach to the other lead.

**Discussion**

No band gap is detected experimentally for graphene on Pt[18] and Au[49]. However, band gaps of 0.18, 0.25, and 0.32 eV are detected for graphene on Cu/Ni[49], Cu[50], and Ag/Ni[49] substrates, respectively, in angle-resolved photoelectron spectroscopy (ARPES). If graphene is identified as single layer one, these measurements are apparently in contradiction with the calculated zero-gap for SLG on Cu and Ag substrates[35,36,38]. One possible solution to such a great discrepancy is to identify graphene as BLG or ABC-stacked TLG, since CVD growth on Ni substrate can yield few-layer graphene with random stacking order[16,17] and on Cu substrate can yield bilayer[12] and few-layer[13,14] graphene, breaking the self-limiting nature of growth process. According to our calculations, BLG on Cu, and Ag (111) surfaces has a band gap of 0.11, 0.13 eV, and ABC-stacked TLG has a band gap of 0.181 and 0.203 eV, respectively, all of which are comparable with the measured values[18,49,50].

The other interesting point is the different strength of interaction in different categories of interfaces. In terms of so-called *d* band model, the bond strength increases when moving to left and up in the transition metal series[51]. As moving from the right to the left, the *d* band moves up in energy, the filling of *d* band decreases and the antibonding graphene-metal *d* states become more depopulated, resulting in a strong bonding. Rises of the 3*d*, 4*d*, and 5*d* states are observed as going from Ni, Co, to Ti, from Ag to Pd, and from Au to Pt, respectively. From Table 1, we indeed have $E_b$(Ti) > $E_b$(Co) > $E_b$(Ni) > $E_b$(Cu) for 3*d* metals, $E_b$(Pd) > $E_b$(Ag) for 4*d* metals and $E_b$(Pt) > $E_b$(Au) for 5*d* metals. The same binding difference is available for SLG on metal substrate[35,36]. Because the Ni and Co 3*d* states of the minority-spin are higher in energy than those of the majority-spin (Figure 5), TLG should interact more strongly with the minority-spin states of Ni/Co in terms of this model.

As moving down in one group, relativistic effects become more remarkable in the core electrons, therefore the *d*-state orbitals of metals diffuse more widely, resulting in a worse overlapping of graphene $\pi$ states and metal *d* states and a weaker binding[23]. Besides, the strength of covalent bond generally decreases with the increase of the atomic radius in one group. The calculated binding energy for group 10 metals (Ni, Pd, and Pt) with TLG indeed follows this rule and we have $E_b$(Ni) > $E_b$(Pd) > $E_b$(Pt) (See Table 1). In fact, Ti, Pd, and Pt are the three represents of the three classes of metals: BLG and TLG are strongly chemisorbed on Ni, and weakly chemisorbed on Pd, while the adsorption of graphene on Pt degenerates into a physisorption. The interaction strength change of graphene with group 9 metals Co, Rh, and Ir also obeys the same rule[15,23-25].

Epitaxial ABA-stacked TLG on Ru (0001) surface has been investigated by Sutter *et al.*[22] The electronic structure determined by selected-area APPES shows BLG-like $\pi$ band dispersion. According to their DFT calculation, the two uncontacted graphene layers on Ru (0001) surface behaves like freestanding BLG without a band gap though they are heavily *n*-type doped (the top of the valence band is located at −0.30 ± 0.05 eV below $E_f$). This result is somewhat surprising because the potential difference induced by graphene-metal electron redistribution will destroy the inversion symmetry of the two uncontacted graphene layers. We therefore recalculated the ABA-stacked TLG/Ru contact using the same parameters set by Sutter *et al.*[22] The calculated electronic structure is shown in Figure S4. As expected, a band gap of 0.127 and 0.147 eV is opened in both the CASTEP and VASP calculations. The top of the valence band is located at −0.267 and −0.261 eV below $E_f$ in the CASTEP and VASP calculations (Supplementary information, Figure S4), respectively, consistent with the micro-ARPES data −0.30 ± 0.05 eV[22]. We note that the DFT calculation of Gong *et al.*[47] for ABA-stacked TLG/Ni contact also found an energy gap of $E_g$ = 0.133 eV for the majority-spin band, comparable with our value of 0.191 eV. This calculation also supports our results that the band gap of the uncontacted two layers is generally opened by the charge redistribution between metal and TLG in the second category of interfaces.

In summary, we present the first systematic first-principles investigation on the interfacial properties of BLG and TLG on a variety of metal substrates. According to the adsorption strength and electronic properties, the BLG/metal and TLG/metal interfacial structures can be classified into three categories. In the first category of interfaces, B(T)LG are physisorbed on Al, Ag, Cu, Au, and Pt substrates; a band gap of 0.1 – 0.2 eV is opened for BLG, and a stacking-sensitive band gap is opened for TLG, with the values of 0 − 0.061 and 0 − 0.249 eV for ABA- and ABC-stacking styles, respectively. In the second category of interfaces, B(T)LG are chemisorbed on Ti, Ni, and Co substrates; the bands of the bottom layer graphene is strongly perturbed, but those of the upper layer graphene of BLG is intact and a stacking-insensitive band gap is opened for the two uncontacted layers of TLG. In the third category of interfaces, B(T)LG are weakly chemisorbed on Pd substrate; a band gap of 0.12 eV is opened for the upper layer graphene of BLG and a band gap of 0.064 and 0.308 eV is opened for the two uncontacted layers of ABA- and ABC-stacked TLG, respectively. An *ab initio* quantum transport simulation is performed for a two-probe model made of BLG contacted with Al or Ti electrodes. A transmission minimum and a transport gap are observed in the transmission spectrum with Al contact. By contrast, there is only one transmission minimum in the transmission spectrum with Ti contact due to the strong binding in the electrodes. This fundamental study not only provides a deeper insight into the interaction between B(T)LG and metal substrates but also helps to B(T)LG-based device study because of inevitable B(T)LG/metal contact.

**Methods**

We use six layers of metal atoms (Ni, Co, Cu, Al, Ag, Cu, Pt, and Au) in (111) orientation and Ti in (0001) orientation to simulate the metal surface, and a hexagonal supercell is constructed with a BLG or TLG adsorbed on one side of the metal surface, as shown in Figure 1. We fix in-plane lattice constant of TLG to the experimental value $a = 2.46$ Å. The $1 \times 1$ unit cells of Ni, Co, and Cu (111) faces are adjusted to graphene $1 \times 1$ unit cell, and $\sqrt{3} \times \sqrt{3}$ unit cells of Ti (0001) face and Al, Ag, Cu, Pt, and Au (111) faces are adjusted to graphene $2 \times 2$ unit cell. The approximation is reasonable since the metal surfaces have a small lattice constants mismatch of less than 4% with that of graphene, as seen in Table 1. A vacuum buffer space of at least 12 Å is set.

The geometry optimizations and electronic structure calculations are performed with the ultrasoft pseudopotentials[52] plane-wave basis set with energy cut-off of 350 eV, implemented in the CASTEP code[53]. Generalized gradient approximation (GGA) of Perdew–Burke–Ernzerhof (PBE) form[54] to the exchange-correlation functional is used. To account for the dispersion interaction between graphene, a DFT-D semiempirical dispersion-correction approach is adopted[55]. During the calculations, the cell shape and the bottom four layers of metal atoms are fixed. To obtain reliable optimized structures, the maximum residual force is less than 0.01 eV/Å and energies are converged to within $5 \times 10^{-6}$ eV per atom. The Monkhorst-Pack[56] $k$-point mesh is sampled with a separation of about 0.02 and 0.01 Å$^{-1}$ in the Brillouin zone, respectively, during the relaxation and electronic calculation periods. The component of the energy band and the plane-averaged excess electron density are analyzed via additional calculations based on the plane-wave basis set with a cut-off energy of 400 eV and the projector-augmented wave (PAW) pseudopotential implemented in the VASP code[57,58]. The electronic structures generated by the two packages are nearly indistinguishable.

TLG/Ru(0001) interface model is constructed from a slab of six layers of Ru with the bottom four layers are fixed and a TLG adsorbed on one side. Following the previous work by Sutter *et al.*[22], ABA-stacked TLG is strained to match the Ru lattice parameter $a = 2.68$ Å. Using the same calculation parameters[22], the ABA-stacked TLG/Ru contact is recalculated by using the CASTEP and VASP codes, respectively. Namely, ultrasoft pseudopotential[52] plane-wave basis set with energy cutoff of 340 eV is used. The local density approximation (LDA) in the Ceperley-Alder form is used for the exchange and correlation functional[59,60]. The Monkhorst-Pack[56] $k$-point is sampled by a $15 \times 15$ mesh in the Brillouin zone.

To study how the metallic contacts affect the transport properties of the BLG devices, a two-probe model made of BLG is built, and the BLG channel is contacted with two Al/Ti electrodes (source and drain). We perform transport calculations at zero source-drain bias by using the DFT method coupled with nonequilibrium Green's function (NEGF) method, which are implemented in ATK 11.2 package[61-63]. Single-zeta (SZ) basis set is used, the real-space mesh cutoff is 150 Ry., and the temperature is set at 300 K. The local-density-approximation (LDA)[59,60] is employed for the exchange–correlation functional. The electronic structures of electrodes and central region are calculated with a Monkhorst–Pack[56] $50 \times 1 \times 100$ and $50 \times 1 \times 1$ $k$-point grid, respectively.

**Acknowledgement** This work was supported by the National Natural Science Foundation of China (Nos. 11274016, 51072007, 91021017, 11047018, and 60890193), the National Basic Research Program of China (Nos. 2013CB932604 and 2012CB619304), Fundamental Research Funds for the Central Universities, National Foundation for Fostering Talents of Basic Science (No. J1030310/No. J1103205), Program for New Century Excellent Talents in University of MOE of China, and Nebraska Research Initiative (No. 4132050400) and DOE DE-EE0003174 in the United States. J. Zheng also acknowledges the financial support from the China Scholarship Council.

**Author Contributions.** The idea was conceived by J. L. The calculation was performed by J. Z. and Y. W. The data analyses were performed by J. Z., Y. W., L. W., R. Q., and J. L.. Z. N., W. M., D. Y., J. S., and Z. G. took part in discussion. This manuscript was written by J. Z., Y. W., and J. L.. All authors reviewed this manuscript.

**Additional Information.**

Competing Financial Interests

The authors declare no competing financial interests.

**Author Information**

● **Affiliations**

**State Key Laboratory of Mesoscopic Physics and Department of Physics, Peking University, Beijing 100871, P. R. China**

Jiaxin Zheng, Yangyang Wang, Ruge Quhe, Zeyuan Ni, Dapeng Yu, Junjie Shi, Zhengxiang Gao, Jing Lu

**Academy for Advanced Interdisciplinary Studies, Peking University, Beijing 100871, P. R. China**

Jiaxin Zheng, Ruge Quhe

**Department of Physics, University of Nebraska at Omaha, Omaha, Nebraska 68182-0266**

Lu Wang, Wai-Ning Mei

● **Corresponding authors**

Correspondence to: Jing Lu (jinglu@pku.edu.cn)

**Table 1.** Calculated interfacial distance $d_{C-M}$, binding energy $E_b$ of per carbon atom, work functions $W$, band gap $E_g$, and Fermi-level shift $\Delta E_f$ for BLG and ABA- and ABC-stacked TLG on various metal surfaces. Because the electronic structure of the contacted layer graphene is strongly perturbed on Ni, Co, Ti, and Pd surfaces, we give the Fermi-level shift and band gap for the uncontaced single layer graphene of BLG and two graphene layers of TLG, respectively. $a_{hex}^{exp}$ is the experimental lattice parameters of the surface unit cells shown in Figure 1b and 1d, and $W_M$ the calculated work functions for various clean metal surfaces.

| Substrates | $a_{hex}^{exp}$ (Å) | $W_M$ (eV) | BLG | | | | | ABA-stacked TLG | | | | | ABC-stacked TLG | | | | |
|---|---|---|---|---|---|---|---|---|---|---|---|---|---|---|---|---|---|
| | | | $d_{C-M}$ (Å) | $E_b$ (eV) | $W$ (eV) | $E_g$ (eV) | $\Delta E_f$ (eV) | $d_{C-M}$ (Å) | $E_b$ (eV) | $W$ (eV) | $E_g$ (eV) | $\Delta E_f$ (eV) | $d_{C-M}$ (Å) | $E_b$ (eV) | $W$ (eV) | $E_g$ (eV) | $\Delta E_f$ (eV) |
| Al | 4.96 | 4.06 | 3.45 | 0.057 | 3.80 | 0.200 | −0.361 | 3.43 | 0.035 | 3.82 | 0.061 | −0.293 | 3.39 | 0.039 | 4.09 | 0.249 | −0.331 |
| Ag | 5.00 | 4.46 | 3.41 | 0.052 | 4.1 | 0.131 | −0.182 | 3.24 | 0.033 | 3.92 | 0.032 | −0.204 | 3.24 | 0.035 | 4.35 | 0.203 | −0.249 |
| Cu | 2.56 | 4.84 | 3.19 | 0.063 | 4.27 | 0.110 | −0.171 | 3.13 | 0.051 | 4.39 | 0.041 | −0.182 | 3.21 | 0.047 | 4.60 | 0.181 | −0.137 |
| Au | 4.99 | 5.17 | 3.46 | 0.051 | 4.89 | 0.102 | 0.230 | 3.44 | 0.032 | 4.71 | 0 | 0.171 | 3.43 | 0.036 | 4.96 | 0 | 0.141 |
| Pt | 4.81 | 5.82 | 3.53 | 0.059 | 5.19 | 0.113 | 0.351 | 3.17 | 0.037 | 5.15 | 0 | 0.204 | 3.13 | 0.039 | 5.42 | 0.018 | 0.145 |
| Ti | 5.11 | 4.20 | 2.18 | 0.21 | 3.89 | 0 | −0.283 | 2.20 | 0.188 | 3.58 | 0.142 | −0.203 | 2.20 | 0.208 | 3.82 | 0.100 | −0.360 |
| Co | 2.51 | 4.97 | 2.17 | 0.099 | 4.33 | 0 | −0.428[a] / −0.381[b] | 2.05 | 0.138 | 4.46 | 0.177[a] / 0.156[b] | −0.248[a] / −0.259[b] | 2.04 | 0.172 | 4.15 | 0.226[a] / 0.160[b] | −0.278[a] / −0.242[b] |
| Ni | 2.49 | 4.95 | 2.34 | 0.094 | 4.32 | 0 | −0.364[a] / −0.280[b] | 2.13 | 0.123 | 4.25 | 0.191[a] / 0.157[b] | −0.238[a] / −0.257[b] | 2.13 | 0.156 | 4.40 | 0.185[a] / 0.229[b] | −0.247[a] / −0.278[b] |
| Pd | 4.76 | 5.33 | 2.70 | 0.083 | 4.74 | 0.124 | −0.160 | 2.54 | 0.085 | 4.88 | 0.064 | −0.104 | 2.50 | 0.103 | 4.82 | 0.308 | −0.220 |

[a] Majority-spin band, [b] Minority-spin band.

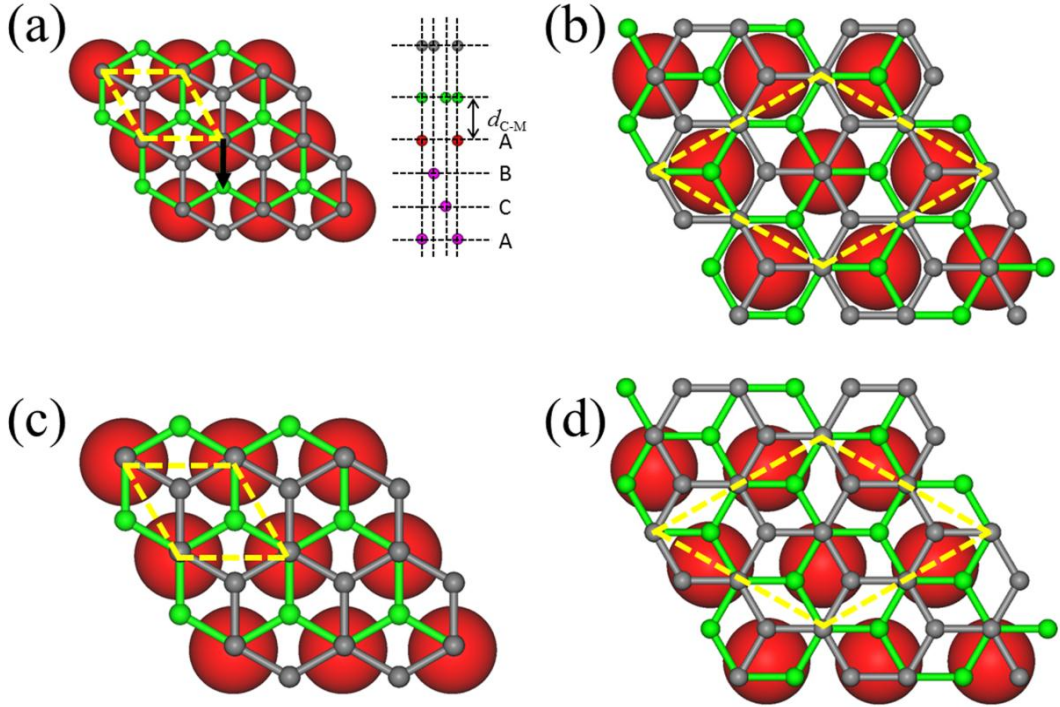

**Figure 1: Interfacial structures of B(T)LG on metal substrates.** (a) Top and side views of the most stable configuration for SLG (the green balls)[35,36] and BLG on Ni, Co, and Cu (111) surfaces. (b) Top views of the most stable configuration for SLG[35,36] and BLG on Ti (0001) surface, and Pd, Al, Ag, Au, and Pt (111) surfaces. (c) and (d) The relaxed configurations for TLG on the corresponding metal substrates with split alignment of the first (innermost) graphene layer to metals compared to (a) and (b). Red and purple balls denote metal atoms of the first and rest layers, respectively. Green and gray balls denote the first and second layers of graphene, respectively. The third (outermost) graphene layer (not shown) is vertically aligned with the first layer for ABA stacking mode and it has a vector translation (labeled by a black arrow) with respect to the first layer for ABC stacking mode. $d_{C-M}$ is the equilibrium distance between the metal surface and the bottom layer graphene. The yellow diamonds represent unit cells.

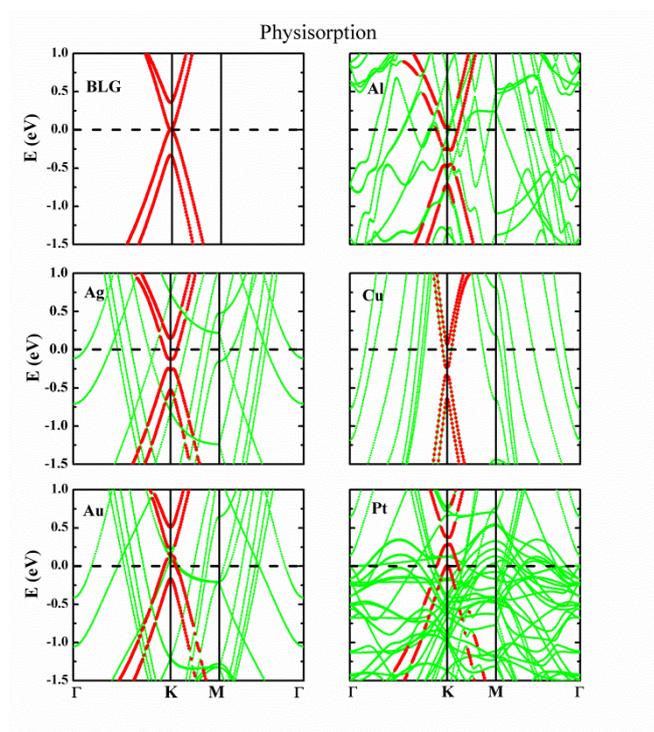

**Figure 2.** Band structures of freestanding BLG and BLG physisorbed on Al, Ag, Cu, Au, and Pt (111) substrates. The Fermi level is set to zero. BLG dominated bands (red) are plotted against the metal projected bands (green).

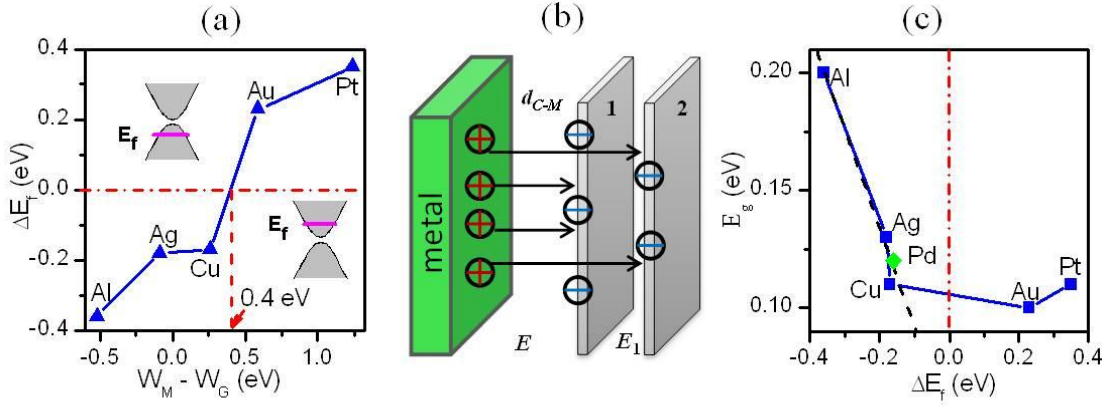

**Figure 3.** (a) Calculated Fermi-level shift as a function of $W_M - W_G$, the difference between the clean metal and graphene work functions. $W_M - W_G = 0.4$ eV is the cross point from *n*- to *p*-type doping. (b) Schematic of the BLG/metal contacts. $E$ and $E_1$ denote the electric fields between metal and graphene and between the graphene layers, respectively. (c) Band gap as a function of $\Delta E_f$ in BLG physisorbed on the metal surfaces. The red dot-dashed line in (c) is a boundary of *n*- and *p*-type doping region. The black dashed line in (c) is a linear fit to the $E_g$ – $\Delta E_f$ data in the *n*-type doped region. The band gap (green diamond) of the upper layer graphene for BLG weakly chemisorbed on Pd surface is also given.

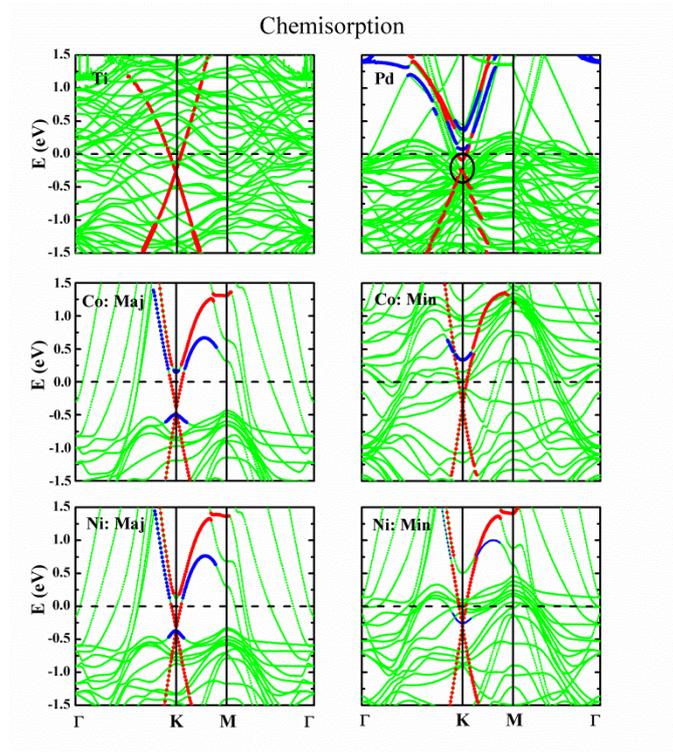

**Figure 4.** Band structures of BLG chemisorbed on Ni, Co, and Pd (111) and Ti (0001) substrates. The Fermi level is set to zero. Green line: metal surface bands; blue line: bands of the upper layer graphene; red line: bands of the bottom layer graphene.

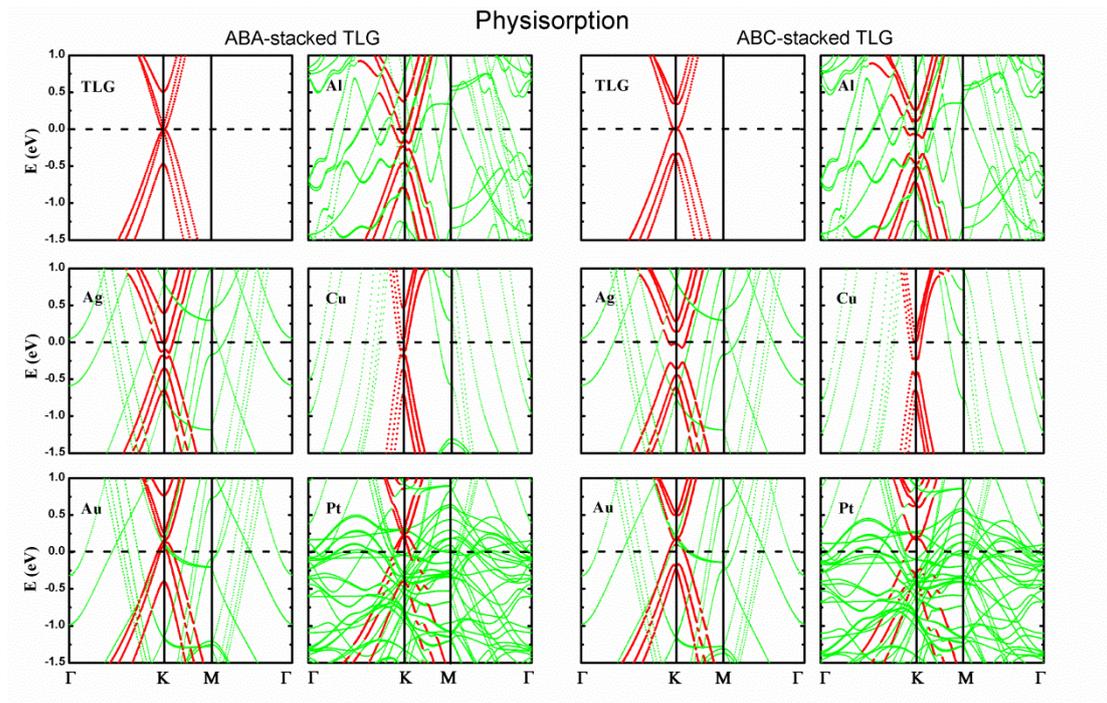

**Figure 5.** Band structures of ABA- and ABC-stacked TLG physisorbed on Al, Ag, Cu, Au, and Pt (111) surfaces. The Fermi level is set at zero. TLG-dominated bands (red) are plotted against the metal projected bands (green). The first and third top panels correspond to the band structure of freestanding ABA- and ABC- stacked TLG with graphene 2×2 supercell, respectively.

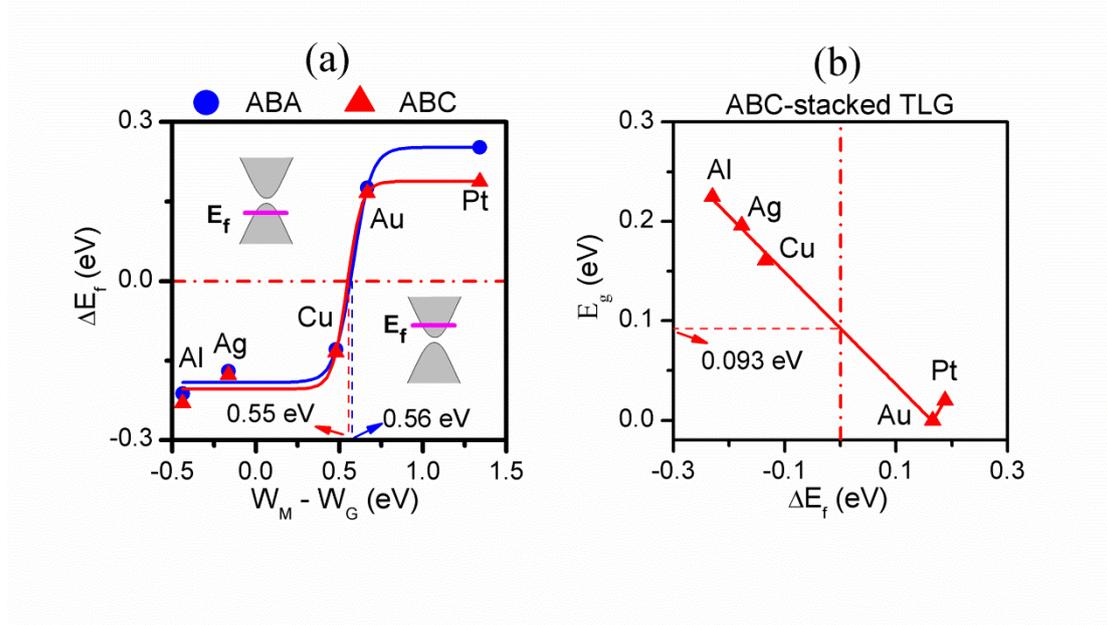

**Figure 6.** Calculated Fermi-level shift $\Delta E_f$ as a function of $W_M - W_G$, the difference between the clean metal and TLG work functions, for (a) ABA- and ABC-stacked TLG physisorbed on the metal surfaces. (b) Band gap $E_g$ as a function of $\Delta E_f$ in ABC-stacked TLG physisorbed on the metal surfaces. The red dash-dot line is the boundary of the *n*- and *p*-type doping region.

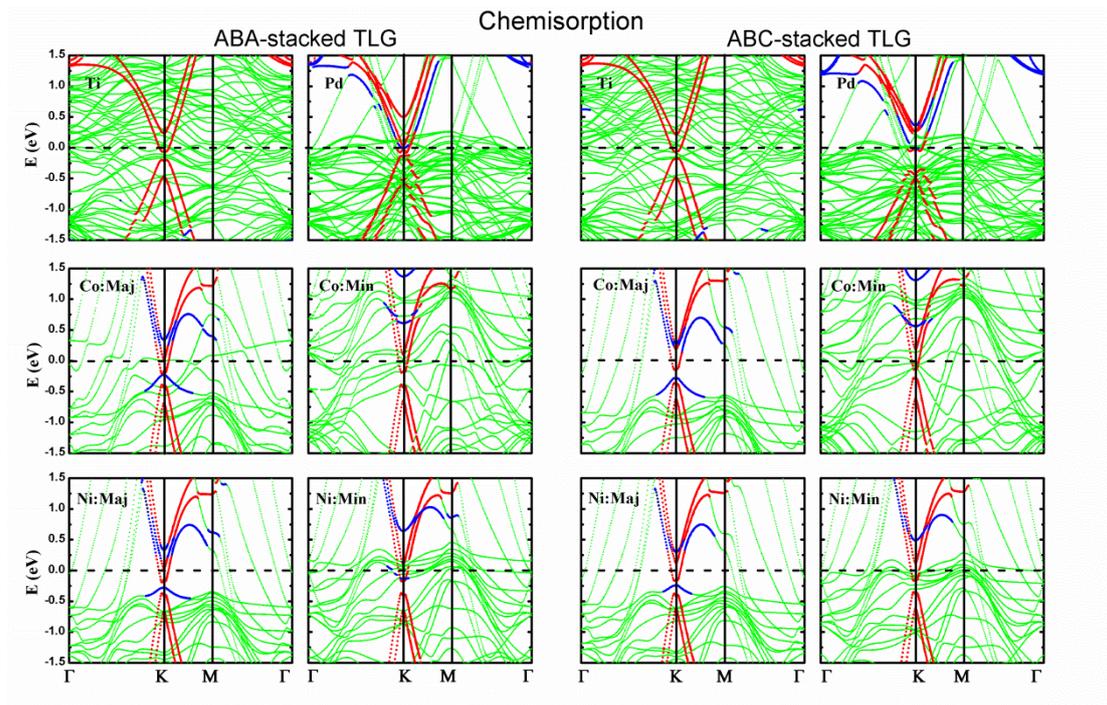

**Figure 7.** Band structures of ABA- and ABC-stacked TLG chemisorbed on Ti (0001), Co, Ni, and Pd (111) surfaces. The Fermi level is set at zero. TLG-dominated bands are plotted against the metal projected bands (green). Blue and red lines depict the bands with weight projected on the innermost graphene layer and the outer graphene bilayer, respectively. The labels Maj/Min represent the majority- and minority-spin bands, respectively.

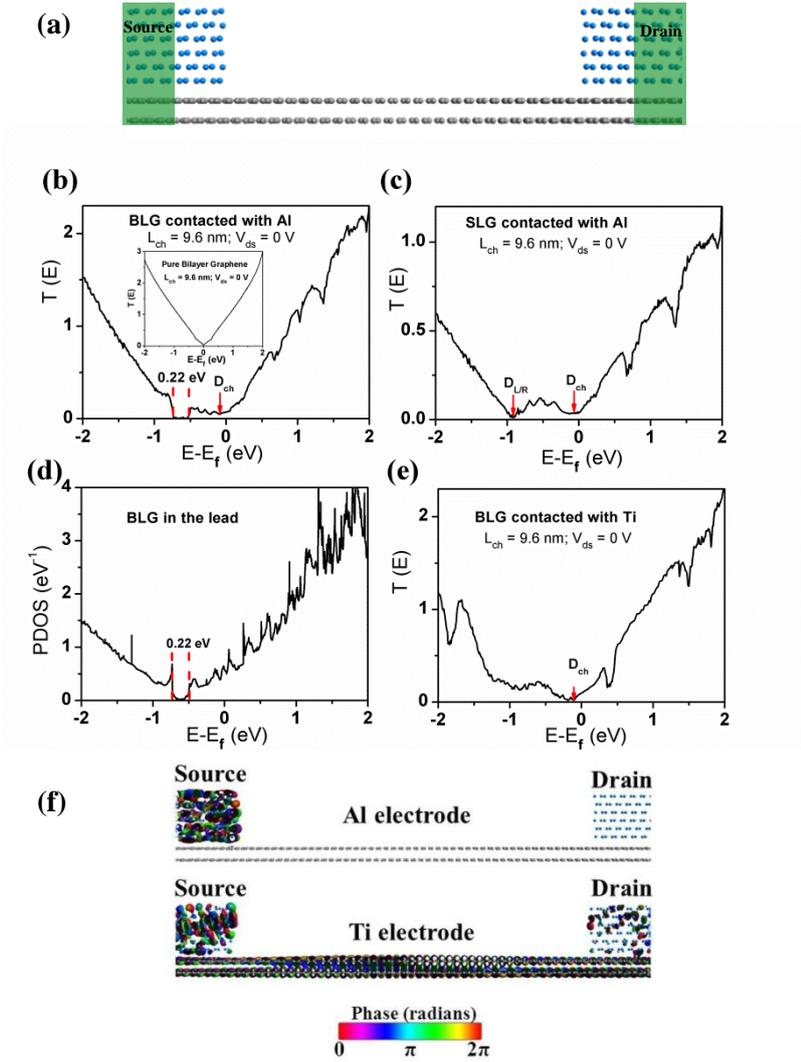

**Figure 8.** (a) Two-probe model. The length of the channel is $L_{ch}$ = 9.6 nm. Gray ball: C; blue ball: Al or Ti. (b) Zero-bias transmission spectrum with Al electrodes. Inset: transmission spectrum of a freestanding BLG with the same $L_{ch}$. (c) Zero-bias transmission spectrum of SLG contacted with Al electrodes with $L_{ch}$ = 9.6 nm. (d) Projected density of states (PDOS) of BLG contacted with Al electrodes. (e) Zero-bias transmission spectrum of BLG contacted with Ti electrodes with $L_{ch}$ = 9.6 nm. (f) Transmission eigenstates at $E - E_f$ = –0.6 eV and at $k$ = ($\pi/3a$, 0) with Al and Ti electrodes, respectively. The isovalue is 0.2 *a.u.*.